\documentclass[final]{aipproc}

\layoutstyle{6x9}

\renewcommand{\bar}[1]{\overline{#1}}
\newcommand{\VEV}[1]{\left\langle{#1}\right\rangle}
\newcommand{\ket}[1]{\,\left|\,{#1}\right\rangle}

\begin{document}

\title{Hadron Spectroscopy and Wavefunctions in QCD and the AdS/CFT Correspondence
}


\author{Stanley J. Brodsky}{
  address={Stanford Linear Accelerator Center, Stanford University, Stanford, California 94309, USA}
}

\author{Guy F. de T\'eramond}{
  address={Universidad de Costa Rica, San Jos\'e, Costa Rica}
}

\begin{abstract}
The AdS/CFT correspondence has led to important insights into the
properties of quantum chromodynamics even though QCD is a broken
conformal theory. We have recently shown how a holographic model
based on a truncated AdS space can be used to obtain the hadronic
spectrum of light $q \bar q, qqq$ and $gg$ bound states.  Specific
hadrons are identified by the correspondence of string modes with
the dimension of the interpolating operator of the hadron's valence
Fock state, including orbital angular momentum excitations. The
predicted mass spectrum is linear $M \propto L$ at high orbital
angular momentum, in contrast to the quadratic dependence $M^2
\propto L$ found in the description of spinning strings. Since only
one parameter, the QCD scale $\Lambda_{QCD}$, is introduced, the
agreement with the pattern of physical states is remarkable. In
particular, the ratio of $\Delta$ to nucleon trajectories is
determined by the ratio of zeros of Bessel functions. The
light-front quantization of gauge theories in light-cone gauge
provides a frame-independent wavefunction representation of
relativistic bound states, simple forms for current matrix elements,
explicit unitarity, and a trivial vacuum. The light-front Fock-state
wavefunctions encode the bound state properties of hadrons in terms
of their quark and gluon degrees of freedom at the amplitude level.
One can also use the extended AdS/CFT space-time theory to obtain a
model for hadronic light-front wavefunctions, thus providing a
relativistic description of hadrons in QCD at the amplitude level.
The model wavefunctions display confinement at large inter-quark
separation and conformal symmetry at short distances. In particular,
the scaling and conformal properties of the LFWFs at high relative
momenta agree with perturbative QCD. These AdS/CFT model
wavefunctions could be used as an initial ansatz for a variational
treatment of the light-front QCD Hamiltonian. We also show how
hadron form factors in both the space-like and time-like regions can
be predicted.

\end{abstract}

\maketitle

\section{Introduction}
A central goal in quantum chromodynamics is to describe the
structure and dynamics of hadrons at the amplitude level. The
light-front Fock expansion provides a physical description of
hadrons as composites of quarks and gluons analogous to the
$\psi(\vec p)$ momentum-space wavefunction description of
nonrelativistic bound states of the Schr\"odinger theory.  The
light-front wavefunctions $ \psi_{n/H}(x_i, \vec k_{\perp i},
\lambda_i )$  are functions of the constituent light-cone fractions
$x_i={k^+_i\over P^+} ={(k^0+k^z)_i\over P^+}$, relative transverse
momenta $\vec k_{\perp i}$, and spin projections $S^z_i=\lambda_i.$
They are relativistic and frame-independent, describing all
particle number excitations $n$ of the hadrons.

The light-front Fock expansion follows from the quantization of QCD
at fixed light-front time $x^+ = x^0 +x^3$. The bound-state hadronic
solutions $\ket{\Psi_H}$ are eigenstates of the light-front
Heisenberg equation $H_{LF} \ket{\Psi_H} = M^2_H
\ket{\Psi_H}$~\cite{Brodsky:1997de}. The spectrum of QCD is given by
the eigenvalues $M^2_H$.  The projection of each hadronic
eigensolution on the free Fock basis: $\VEV{n\,|\,\Psi_H} \equiv
\psi_{n/H}(x_i, \vec k_{\perp i}, \lambda_i )$ then defines the LF
Fock expansion in terms of the quark and transversely polarized
gluon constituents in $A^+=0$ light-cone gauge.  The expansion has
only transversely polarized gluons. The freedom to choose the
light-like quantization four-vector provides an explicitly covariant
formulation of light-front quantization and can be used to determine
the analytic structure of light-front wave functions and to define a
kinematical definition of angular momentum~\cite{Brodsky:2003pw}.
The front form thus provides a consistent definition of relative
orbital angular momentum and $J^z$ conservation: the total spin
projection $J^z = \sum^n_{i=1} S^z_i + \sum^{n-1}_i L^z_i$ is
conserved in each Fock state. The cluster decomposition
theorem~\cite{Brodsky:1985gs} and the vanishing of the ``anomalous
gravitomagnetic moment" $B(0)$~\cite{Teryaev:1999su} are immediate
properties of the LF Fock wavefunctions~\cite{Brodsky:2000ii}.

Given the light-front wavefunctions $\psi_{n/H}(x_i, \vec k_{\perp
i}, \lambda_i )$, one can compute a large range of hadron
observables. For example, the valence and sea quark and gluon
distributions which are measured in deep inelastic lepton scattering
are defined from the squares of the LFWFS summed over all Fock
states $n$. Form factors, exclusive weak transition
amplitudes~\cite{Brodsky:1998hn} such as $B\to \ell \nu \pi$. and
the generalized parton distributions~\cite{Brodsky:2000xy} measured
in deeply virtual Compton scattering are (assuming the ``handbag"
approximation) overlaps of the initial and final LFWFS with $n
=n^\prime$ and $n =n^\prime+2$. The gauge-invariant distribution
amplitude $\phi_H(x_i,Q)$ defined from the integral over the
transverse momenta $\vec k^2_{\perp i} \le Q^2$ of the valence
(smallest $n$) Fock state provides a fundamental measure of the
hadron at the amplitude level~\cite{Lepage:1979zb,Efremov:1979qk};
they  are the nonperturbative input to the factorized form of hard
exclusive amplitudes and exclusive heavy hadron decays in
perturbative QCD. The resulting distributions obey the DGLAP and
ERBL evolution equations as a function of the maximal invariant
mass, thus providing a physical factorization
scheme~\cite{Lepage:1980fj}. In each case, the derived quantities
satisfy the appropriate operator product expansions, sum rules, and
evolution equations.  However, at large $x$ where the struck quark
is far-off shell, DGLAP evolution is quenched~\cite{Brodsky:1979qm},
so that the fall-off of the DIS cross sections in $Q^2$ satisfies
inclusive-exclusive duality at fixed $W^2.$

The light-front Fock-state wavefunctions encode the bound state
properties of hadrons in terms of their quark and gluon degrees of
freedom at the amplitude level. They display  novel features, such
as intrinsic gluons, asymmetric sea-quark distributions $\bar u(x)
\ne \bar d(x)$, $\bar s(x) \ne s(x)$, and intrinsic heavy-quark Fock
states~\cite{Brodsky:1980pb}.  Intrinsic charm and bottom quarks
appear at large $x$ in the light-front wavefunctions since this
minimizes the invariant mass and off-shellness of the higher Fock
state. One can use the operator product expansion to show that the
probability of such states scales as $1/M^2_Q$ in contrast to
$1/M_\ell^4$ fall-off of abelian theory~\cite{Franz:2000ee}. The
remarkable observations of the SELEX experiment of the  double-charm
baryon $\Xi_{ccd}$ in $p A \to \Xi_{ccd} X$ and $\Sigma^- A \to
\Xi_{ccd} X$ at  large $x_F$~\cite{Engelfried:2005kd} provides
compelling evidence for double-charm intrinsic Fock states in the
proton.  The coherence of multi-particle correlations within the
Fock states leads to higher-twist bosonic processes such as $e (qq)
\to e^\prime (qq)^\prime$; although suppressed by inverse powers of
$Q^2$, such subprocesses are important in the duality regime of
fixed $W^2$, particularly in $\sigma_L$~\cite{Brodsky:1982qn}. In
the case of nuclei, one must include non-nucleonic ``hidden color"
~\cite{Brodsky:1983vf} degrees of freedom of the deuteron LFWF.

\subsection{Measurements of the LFWFs}
The E791 experiments at Fermilab~\cite{Ashery:2002ri,Aitala:2000hb}
has shown how one can measure the valence LFWF directly from the
diffractive di-jet dissociation of a high energy pion $\pi A \to q
\bar q A^\prime$ into two jets, nearly balancing in transverse
momentum, leaving the nucleus intact.  The measured pion
distribution in $x$ and $(1-x)$ is similar to the form of the
asymptotic distribution amplitude and the AdS/CFT prediction
discussed below. The E791 experiment also find that the nuclear
amplitude is  additive in the number of nucleons when the quark jets
are produced at high $k_\perp$, thus giving a dramatic confirmation
of ``color transparency",  a fundamental manifestation of the gauge
nature of QCD~\cite{Bertsch:1981py,Brodsky:1988xz}.

\subsection{Effects of Final State Interactions}

The phase structure of hadron matrix elements is an essential
feature of hadron dynamics. Although the LFWFs are real for a stable
hadron, they acquire phases from initial state and final state
interactions.  In addition, the violation of $CP$ invariance leads
to a specific phase structure of the LFWFs~\cite{bgh}.

Contrary to parton model expectations,  the rescattering of the
quarks in the final state in DIS has  important phenomenological
consequences, such as leading-twist diffractive 
DIS~\cite{Brodsky:2002ue} and the Sivers single-spin
asymmetry~\cite{Brodsky:2002cx}. The Sivers asymmetry depends on the
same matrix elements which produce the anomalous magnetic moment of
the target nucleon as well as the phase difference of the
final-state interactions in different partial waves. The
rescattering of the struck parton generates dominantly imaginary
diffractive amplitudes, giving rise to an effective ``hard pomeron"
exchange and a rapidity gap between the target and diffractive
system, while leaving the target intact. This Bjorken-scaling
physics, which is associated with the Wilson line connecting the
currents in the virtual Compton amplitude survives even in
light-cone gauge. Thus there are contributions to the DIS structure
functions which are not included in the light-front wave functions
computed in isolation and cannot be interpreted as parton
probabilities~\cite{Brodsky:2002ue}.  Diffractive deep inelastic
scattering  in turn leads to nuclear shadowing at leading twist as a
result of the destructive interference of multi-step processes
within the nucleus. In addition, multi-step processes involving
Reggeon exchange leads to antishadowing. In fact, because Reggeon
couplings are flavor specific, antishadowing is predicted to be
non-universal, depending on the type of current and even the
polarization of the probes in nuclear DIS~\cite{Brodsky:2004qa}.

Another interesting consequence of QCD at the amplitude
level is the $Q^2$-independent ``$J=0$ fixed-pole'' contribution
$M(\gamma^* p \to \gamma p) \sim s^0 F(t)$  to the real part of the
Compton amplitude, reflecting the effective contact interaction of
the transverse currents~\cite{Brodsky:1971zh}.  Deeply virtual
Compton scattering can also be studied in the timelike domain from
$e^+ e^- \to H^+ H^- \gamma$; the lepton charge asymmetry and
single-spin asymmetries allow measurements of the relative phase of
timelike form factors and the $\gamma^* \to H^+ H^- \gamma$
amplitude.

\subsection{Nonperturbative Methods for Computing LFWFs}
In principle, one can solve for the LFWFs directly from the
fundamental theory using methods such as discretized light-front
quantization (DLCQ)~\cite{Pauli:1985ps}, the transverse
lattice~\cite{Bardeen:1979xx,Dalley:2004rq,Burkardt:2001jg}, lattice
gauge theory moments~\cite{DelDebbio:1999mq}, Dyson-Schwinger
techniques~\cite{Maris:2003vk}, and Bethe--Salpeter
techniques~\cite{Brodsky:2003pw}. DLCQ has been remarkably
successful in determining the entire spectrum and corresponding
LFWFs in one space-one time field theories~\cite{Gross:1997mx},
including QCD(1+1)~\cite{Hornbostel:1988fb} and supersymmetric
QCD(1+1)~\cite{Harada:2004ck}. The DLCQ boundary conditions allow a
truncation of the Fock space to finite dimensions while retaining
the kinematic boost and Lorentz invariance of light-front
quantization. There are also light-front solutions for Yukawa theory
in physical (3+1) space-time
dimensions~\cite{Brodsky:2002tp,Brodsky:2005yu} with a limited Fock
space. As emphasized by Weinstein and Vary, new effective operator
methods~\cite{Weinstein:2004nr,Zhan:2004ct} which have been
developed for Hamiltonian theories in  condensed matter and nuclear
physics, could also be applied advantageously to light-front
Hamiltonian.  A review of nonperturbative light-front methods may
be found in reference~\cite{Brodsky:2004er}.

As we discuss below, one can use the AdS/CFT correspondence to
obtain a model for hadronic light-front wavefunctions which display
confinement at large inter-quark separation and conformal symmetry
at short distances. In particular, the scaling and conformal
properties of the LFWFs at high relative momenta agree with
perturbative QCD. These AdS/CFT model wavefunctions could be used as
an initial ansatz for a variational treatment of the light-front QCD
Hamiltonian.

\section{AdS/CFT Predictions for Hadron Spectra and Wavefunctions}

The AdS/CFT correspondence~\cite{Maldacena:1997re}, between
strongly-coupled conformal gauge theory and weakly-coupled string
theory in the 10-dimensional $AdS_5 \times S^5$ space is now
providing a remarkable new insight into the hadron wavefunctions of
QCD. The central mathematical principle underlying AdS/CFT duality
is the fact that the group $SO(2,4)$ of Poincar\'e and conformal
transformations of physical $3+1$ space-time has an elegant
mathematical representation on ${\rm AdS}_5$ space where the fifth
dimension has the anti-de Sitter warped metric.  The group of
conformal transformations $SO(2,4)$ in 3+1 space  is isomorphic to
the group of isometries of AdS space, $x^\mu \to \lambda x^\mu$, $r
\to r/\lambda$, where $r$ represents the coordinate in the fifth
dimension. The dynamics at $x^2 \to 0$ in 3+1 space thus matches the
behavior of the theory at the boundary $r \to \infty.$ This allows
one to map the physics of quantum field theories with conformal
symmetry to an equivalent description in which scale transformations
have an explicit representation in AdS space.

Even though quantum chromodynamics  is a broken conformal theory,
the AdS/CFT correspondence  has led to important insights into the
properties of QCD. For example, as shown by  Polchinski and
Strassler~\cite{Polchinski:2001tt}, the AdS/CFT duality, modified to
give a mass scale, provides a nonperturbative derivation of the
empirically successful dimensional counting
rules~\cite{Brodsky:1973kr,Matveev:1973ra} for the leading power-law
fall-off of  the hard exclusive scattering amplitudes of the bound
states of the gauge theory. The modified theory generates the hard
behavior expected from QCD instead of the soft behavior
characteristic of strings. Other important applications include the
description  of spacelike hadron form factors at large transverse
momentum~\cite{Polchinski:2001ju} and deep inelastic scattering
structure functions at small $x$~\cite{Polchinski:2002jw}. The
power falloff of hadronic light-front wave functions (LFWF)
including states with nonzero orbital angular momentum is also
predicted~\cite{Brodsky:2003px}.

In the original formulation by Maldacena~\cite{Maldacena:1997re}, a
correspondence was established between a  supergravity  string
theory on a curved background and a conformally invariant
$\mathcal{N} = 4$ super Yang-Mills theory in four-dimensional
space-time. The higher dimensional theory is $AdS_5 \times S^5$
where $R = ({4 \pi g_s N_C})^{1/4} \alpha_s'^{1/2}$ is the radius of
AdS and the radius of the five-sphere and $\alpha_s'^{1/2}$ is the
string scale. The extra dimensions of the five-dimensional sphere
$S^5$ correspond to the $SU(4) \sim SO(6)$ global symmetry which
rotates the particles present in the supersymmetric Yang Mills
supermultiplet in the adjoint representation of $SU(N_C)$. In our
application to QCD, baryon number in QCD is represented as a Casimir
constant on $S^5.$

The reason why AdS/CFT duality can have at least approximate
applicability to physical QCD is based on the fact that the
underlying classical QCD Lagrangian with massless quarks is
scale-invariant~\cite{Parisi:1972zy}. One can thus take conformal
symmetry as an initial approximation to QCD, and then systematically
correct for its nonzero $\beta$ function and quark
masses~\cite{Brodsky:1985ve}.  This ``conformal template"  approach
underlies the Banks-Zak method~\cite{Banks:1981nn} for expansions of
QCD expressions near the conformal limit and the BLM
method~\cite{Brodsky:1982gc} for setting the renormalization scale in
perturbative QCD applications. In the BLM method the corrections to
a perturbative series from the $\beta$-function are systematically
absorbed into the scale of the QCD running coupling. An important
example is the ``Generalized Crewther Relation"~\cite{Brodsky:1995tb}
which relates the Bjorken and Gross-Llewellyn sum rules at the deep
inelastic scale $Q^2$ to the $e^+ e^-$ annihilation cross sections
at specific commensurate scales $ s^*(Q^2) \simeq 0.52~ Q^2$. The
Crewther relation~\cite{Crewther:1972kn} was originally derived in
conformal theory; however, after BLM scale setting, it becomes a
fundamental test of physical QCD, with no uncertainties from the
choice of renormalization scale or scheme.

QCD is nearly conformal at large momentum transfers where asymptotic
freedom is applicable. Nevertheless, it is remarkable that
dimensional scaling for exclusive processes  is observed even at
relatively low momentum transfer where gluon exchanges involve
relatively soft momenta~\cite{deTeramond:2005kp}.  The observed
scaling of hadron scattering  at moderate momentum
transfers can be understood if the QCD coupling has an infrared
fixed point~\cite{Brodsky:2004qb}. In this sense, QCD resembles a
strongly-coupled conformal theory.

\subsection{Deriving Hadron Spectra from AdS/CFT}

The duality between a gravity theory on $AdS_{d+1}$ space and a
conformal gauge theory at its $d$-dimensional boundary requires one
to match the partition functions at the $AdS$ boundary, $z = R^2/r
\to 0$. The physical string modes $\Phi(x,r) \sim e^{-i P \cdot x}
f(r)$, are plane waves along the Poincar\'e coordinates with
four-momentum  $P^\mu$ and hadronic invariant mass states $P_\mu
P^\mu = \mathcal{M}^2$. For large-$r$ or small-$z$, $f(r) \sim
r^{-\Delta}$, where the dimension $\Delta$ of the string mode must
be the same dimension as that of the interpolating operator
{\small$\mathcal{O}$} which creates a specific  hadron out of the
vacuum: $\langle P \vert \mathcal{O} \vert 0 \rangle  \neq 0$.

The physics of color confinement in QCD can be described in the
AdS/CFT approach by truncating the AdS space to the domain $r_0 < r
< \infty$,  where $r_0 = \Lambda_{\rm QCD} R^2$.   The cutoff at
$r_0$ is dual to the introduction of a mass gap $\Lambda_{\rm QCD}$;
it breaks conformal invariance and is responsible for the generation
of a spectrum of color-singlet hadronic states. The truncation of
the AdS space insures that the distance between the colored quarks
and gluons as they stream into the fifth dimension is limited to $z
< z_0 = {1/\Lambda_{\rm QCD}}$. The resulting $3+1$ theory has both
color confinement at long distances and conformal behavior at short
distances.  The latter property allows one to derive dimensional
counting rules for form factors  and other hard exclusive processes
at high momentum transfer. This approach, which can be described as
a ``bottom-up" approach, has been successful in obtaining general
properties of the low-lying hadron spectra, chiral symmetry
breaking, and hadron couplings in AdS/QCD~\cite{deTeramond:2004qd}
in addition to the hard scattering
predictions~\cite{Polchinski:2001tt,Polchinski:2002jw,Brodsky:2003px}.

In this ``classical holographic model", the quarks and gluons
propagate into the truncated AdS interior according to the AdS
metric without interactions. In effect, their Wilson lines, which
are represented by open strings in the fifth dimension, are rigid.
The resulting equations for spin 0, $\frac{1}{2}$, 1 and
$\frac{3}{2}$ hadrons on $AdS_5 \times S^5$  lead to color-singlet
states with dimension $3, 4$  and $\frac{9}{2}$. Consequently, only
the hadronic states  (dimension-$3$) $J^P=0^-,1^-$ $q \bar q$
pseudoscalar and vector mesons, the (dimension-$\frac{9}{2}$)
$J^P=\frac{1}{2}^+, \frac{3}{2}^+$ $qqq$ baryons, and the
(dimension-$4$) $J^P= 0^+$ $gg$ gluonium states, can be derived in
the classical holographic limit~\cite{deTeramond:2005su}.  This
description corresponds to the valence Fock state as represented by
the light-front Fock expansion. Hadrons also fluctuate in particle
number, in their color representations (such as the hidden-color
states~\cite{Brodsky:1983vf} of the deuteron), as well as in
internal orbital angular momentum. The  higher Fock components of
the hadrons are manifestations of the quantum fluctuations of QCD;
these correspond to the fluctuations of the bulk geometry about the
fixed AdS metric. For spinning strings orbital excitations of
hadronic states correspond to quantum fluctuations about the AdS
metric~\cite{Gubser:2002tv}. It is thus also natural to identify
higher-spin hadrons with the fluctuations around the spin 0,
$\frac{1}{2}$, 1 and $\frac{3}{2}$ classical string solutions of the
$AdS_5$ sector~\cite{deTeramond:2005su}.

\begin{figure}[tbh]
\includegraphics[width=5.5in]{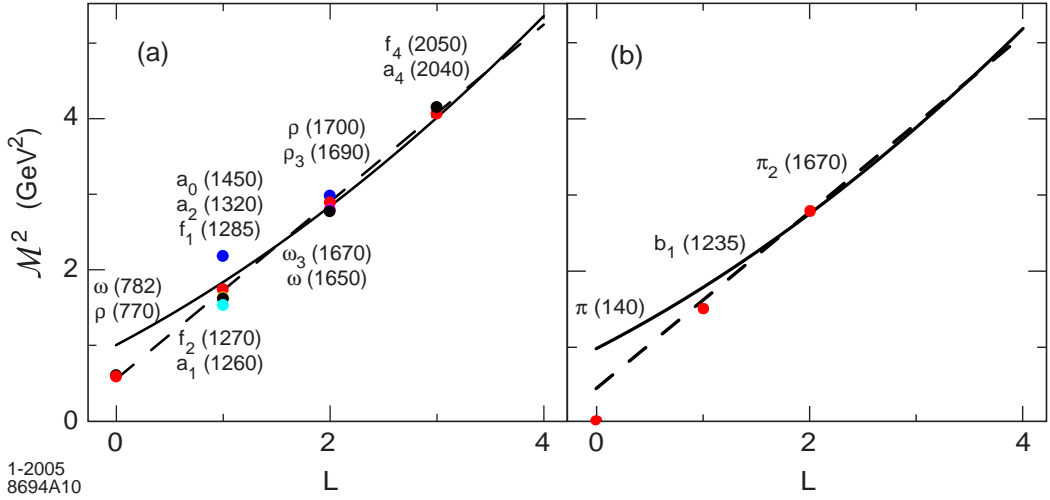}
\caption{Light meson orbital states for $\Lambda_{\rm QCD} = 0.263$
GeV: (a) vector mesons and (b) pseudoscalar mesons. The dashed line
is a linear Regge trajectory with slope 1.16 ${\rm GeV}^2$.}
\label{fig:MesonSpec}
\end{figure}

\begin{figure}[tbh]
\includegraphics[width=5.65in]{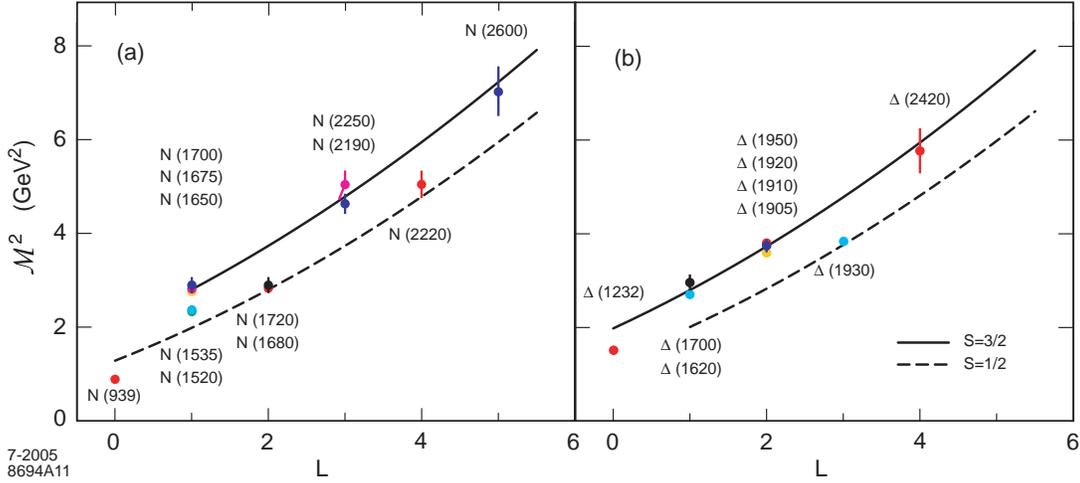}
\caption{Light baryon orbital spectrum for
 $\Lambda_{\rm QCD}$ = 0.22 GeV: (a) nucleons and (b) $\Delta$ states.}
\label{fig:BaryonSpec}
\end{figure}

As a specific example, consider the twist-two (dimension minus spin)
gluonium interpolating operator $\mathcal{O}_{4 + L}^{\ell_1 \cdots
\ell_m} = F D_{\{\ell_1} \dots D_{\ell_m\}} F$ with total internal
space-time orbital momentum $L = \sum_{i=1}^m \ell_i$ and conformal
dimension $\Delta_L = 4 + L$. We match the large $r$ asymptotic
behavior of each string mode to the corresponding conformal
dimension of the boundary operators of each hadronic state while
maintaining conformal invariance. In the conformal limit, an $L$
quantum, which is identified with a quantum fluctuation about the
AdS geometry, corresponds to an effective five-dimensional mass
$\mu$ in the bulk side.  The allowed values of $\mu$ are uniquely
determined by requiring that asymptotically the dimensions become
spaced by integers, according to the spectral relation $(\mu R)^2 =
\Delta_L(\Delta_L - 4)$~\cite{deTeramond:2005su}.  The
four-dimensional mass spectrum follows from the Dirichlet boundary
condition  $\Phi(x,z_o) = 0$, $z_0 = 1 / \Lambda_{\rm QCD}$, on the
AdS string amplitudes for  each  wave functions with  spin $<$ 2.
The eigenspectrum is then determined from the zeros of Bessel
functions, $\beta_{\alpha,k}$. The predicted spectra
\cite{deTeramond:2005su}  of mesons and baryons with zero mass
quarks is shown in Figs.~\ref{fig:MesonSpec} and
\ref{fig:BaryonSpec}. The only parameter is $\Lambda_{\rm QCD} =
0.263$ GeV, and $0.22$ GeV for mesons and baryons, respectively.

\subsection{Dynamics from AdS/CFT}

\begin{figure}[htb]
\includegraphics{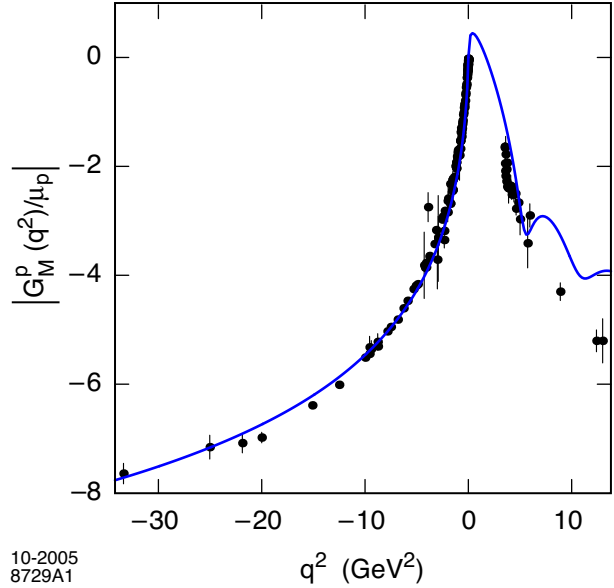}
\caption{Space-like and time-like structure of the proton magnetic
form factor in AdS/QCD for $\Lambda_{QCD} = 0.155$ GeV. The data are
from the compilation given in ref.~\cite{Baldini:1998qn}.  The
prediction in the  domain $0 < q^2 < 4M_p^2$  represents an analytic
continuation into the unphysical region. The
results should be modified for values $Q \sim \Lambda_{QCD}$, where
the simple form for the cavity current $J(Q,z) = z Q K_1(z Q)$ is
not valid.} \label{fig:protonFF}
\end{figure}

Current matrix elements in AdS/QCD are computed from the overlap of
the normalizable modes dual to the incoming and outgoing hadron
$\Phi_I$ and $\Phi_F$ and the non-normalizable mode $J(Q,z)$, dual
to the external source
\begin{equation}
F(Q^2)_{I \to F} \simeq R^{3 + 2\sigma} \int_0^{z_o} \frac{dz}{z^{3
+ 2\sigma}}~
 \Phi_F(z)~J(Q,z)~\Phi_I(z),
\label{eq:FF}
\end{equation}
where $\sigma_n = \sum_{i=1}^n \sigma_i$ is the spin of the
interpolating operator $\mathcal{O}_n$, which creates an $n$-Fock
state $\vert n \rangle$ at the AdS boundary. $J(Q,z)$  has the value
1 at zero momentum transfer as the  boundary limit of the external
current; thus $A^\mu(x,z) = \epsilon^\mu e^{i Q \cdot x} J(Q,z)$.
The solution to the AdS wave equation subject to  boundary
conditions at $Q = 0$ and $z \to 0$ is~\cite{Polchinski:2002jw}
$J(Q,z) = z Q K_1(z Q)$. At large enough $Q \sim r/R^2$, the
important contribution to (\ref{eq:FF}) is from the region near $z
\sim 1/Q$. At small $z$, the $n$-mode $\Phi^{(n)}$ scales as
$\Phi^{(n)} \sim z^{\Delta_n}$, and we recover the power law
scaling~\cite{Brodsky:1973kr},  $F(Q^2) \to \left[1/Q^2\right]^{\tau
- 1}$, where the twist $\tau = \Delta_n - \sigma_n$, is equal to the
number of partons, $\tau_n = n$. A numerical computation for the
proton magnetic form factor in the space and time-like regions, for
the model described here, gives the predictions shown in
Fig.~\ref{fig:protonFF}. The results correspond to a $L=0$ proton
state.   It is interesting to compare the holographic predictions with
a model-independent analysis of nucleon form factors using
dispersion relations~\cite{Baldini:1998qn}.

\section{AdS/CFT Predictions for Light-Front Wavefunctions}

The AdS/QCD correspondence provides a simple description of hadrons
at the amplitude level by mapping string modes to the impact space
representation of LFWFs. It is useful to define the partonic
variables $ x_i \vec r_{\perp i} = x_i \vec R_\perp + \vec b_{\perp i}$,
where $\vec r_{\perp i}$ are the physical position coordinates,
$\vec b_{\perp i}$  are frame-independent internal coordinates,
$\sum_i \vec b_{\perp i} = 0$,  and $\vec R_\perp$ is the hadron
transverse center of momentum $\vec R_\perp = \sum_i x_i \vec
r_{\perp i}$, $\sum_i x_i = 1$.   We find for a two-parton LFWF the
Lorentz-invariant form
\begin{equation}
\widetilde{\psi}_L(x, \vec b_{\perp}) = C ~x(1-x)
~\frac{J_{1+L}\left(\vert\vec b_\perp\vert
\sqrt{x(1-x)}~\beta_{1+L,k} \Lambda_{\rm QCD} \right)} {\vert\vec
b_\perp\vert \sqrt{x(1-x)}}. \label{eq:LFWFbM}
\end{equation}
The $ \beta_{1+L,k}$ are the zeroes of the Bessel functions
reflecting the Dirichlet boundary condition. The variable
$\zeta=\vert\vec b_\perp\vert \sqrt{x(1-x)},$ $0 \leq \zeta \leq
\Lambda^{-1}_{QCD}$, represents the invariant separation between
quarks. In the case of a two-parton state,  it gives a direct
relation between the scale of the invariant separation between
quarks, $\zeta$, and the holographic coordinate in AdS space: $\zeta
= z = R^2/r$. The ground state and first orbital eigenmode are
depicted in the figure below.
\begin{figure}[tbh]
\includegraphics[width=2.9in]{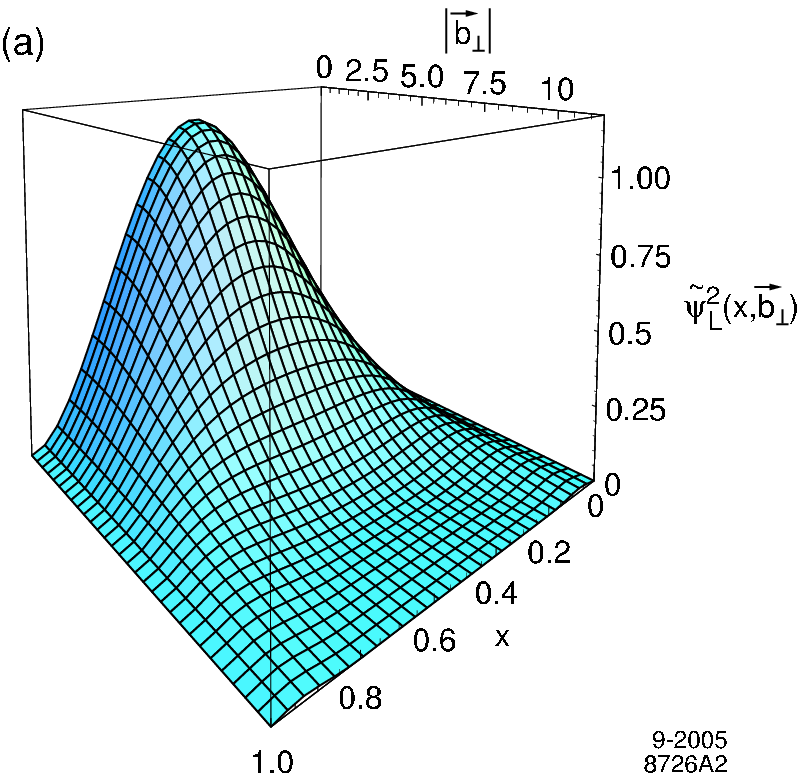}\hfill
\includegraphics[width=2.9in]{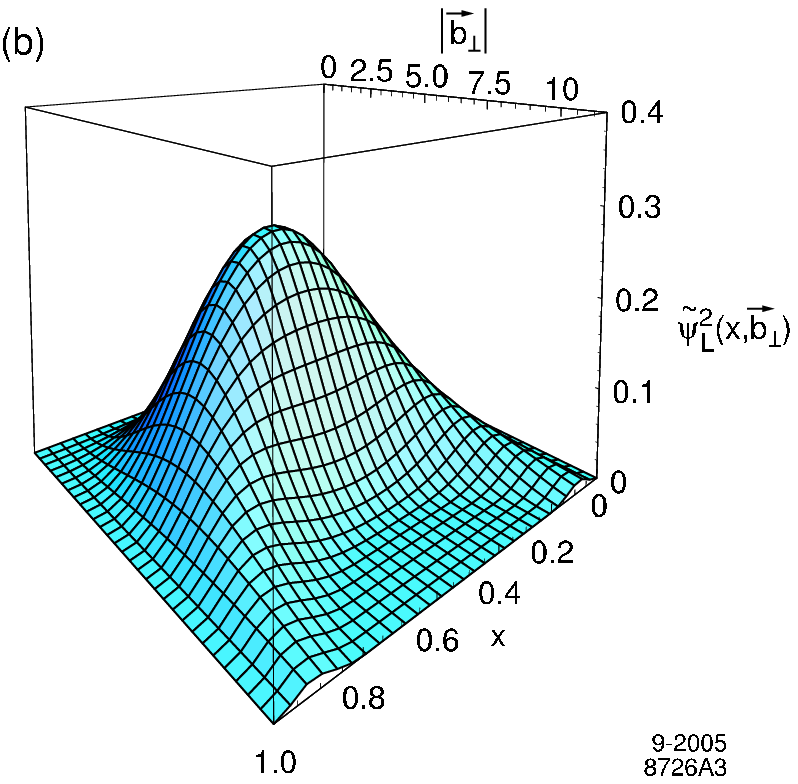}
\caption{Prediction for the square of the two-parton bound-state
light-front wave function $\widetilde{\psi}_L(x,\vec b_\perp)$ as
function of the constituents longitudinal momentum fraction $x$ and
$1-x$ and the impact space relative coordinate $\vec b_\perp$: (a)
$L=0$ and (b) $L=1$.} \label{fig:LFWFb}
\end{figure}
The
distribution in $x$ and $(1-x)$ measured in the E791
experiment  for  diffractive dijet production  $\pi A \to {\rm Jet ~Jet} ~A $ is consistent with
the AdS/CFT prediction~\cite{Ashery:2002ri,Aitala:2000hb}.

\section{Concluding Remarks}

The holographic model is quite successful in describing the known
light hadron spectrum and hadronic form factors. Since only one parameter, the QCD scale
$\Lambda_{QCD}$, is introduced, the agreement with the pattern of
masses of the physical hadronic states and the space and time-like proton
form factor data is remarkable. In particular,
the ratio of $\Delta$ to nucleon trajectories is determined by the
ratio of zeros of Bessel functions.  As we have described, non-zero
orbital angular momentum and higher Fock-states require the
introduction of a Casimir operator derived from quantum
fluctuations. It is interesting to note that the predicted mass
spectrum $M \propto L$ at high orbital angular momentum, in contrast
to the quadratic dependence $M^2 \propto L$ found in traditional
string theory. The only mass scale is $\Lambda_{QCD}$. Only
dimension-$3, \frac{9}{2}$ and 4 states $\bar q q$, $q q q$, and  $g
g$ appear in the duality at the classical level, thus explaining the
suppression of $C=+$ odderon exchange.

We have also shown how one can use the extended AdS/CFT space-time
theory to obtain a model for the form of hadron LFWFs. The  model
wavefunctions display confinement at large inter-quark separation
and conformal symmetry at short distances. In particular, the
scaling and conformal properties of the LFWFs at high relative
momenta agree with perturbative QCD~\cite{Ji:2003fw}. These AdS/CFT
model wavefunctions could be used as an initial ansatz for a
variational treatment of the light-front QCD Hamiltonian.  The
dominance of the quark-interchange mechanism in hard exclusive
processes also emerges naturally from the classical duality of the
holographic model.

\begin{theacknowledgments}
Presented by SJB at the 11th International Conference On Hadron
Spectroscopy, HADRON05, 21-26 Aug 2005, Rio de Janeiro, Brazil. This
work was supported by the Department of Energy contract
DE--AC02--76SF00515.
\end{theacknowledgments}

\end{document}